\documentclass{aa}
\usepackage{graphicx}

\usepackage{txfonts}
\def\be{\begin{equation}}
\def\ee{\end{equation}}
\begin{document}

   \title{Newtonian and general relativistic contribution of gravity
to surface tension of strange stars.}
\titlerunning{surface tension of strange stars ...}
  \author{ Manjari Bagchi \inst{1,2} \and Monika Sinha \inst{1,3}
\and Mira Dey \inst{1,2} \and Jishnu Dey \inst{1,2} \and
Siddhartha Bhowmick \inst{2,5}}
\authorrunning{Bagchi et al}

\offprints{M. Bagchi, email : mnj2003@vsnl.net}

\institute{Dept. of Physics, Presidency College, 86/1 College
Street, Kolkata 700 073, India; Visitor (2005), IUCAA, Pune and
HRI, Allahabad, India., \\ \and Work supported in part by DST
grant no. SP/S2/K-03/2001, Govt. of India. email : kamal1@vsnl.com
\\ \and CSIR NET Fellow,  Govt. of India. \\ \and Department of Physics, Barasat
Govt. College, Barasat, North 24 Parganas, W. Bengal, India.  }

\date{}

\abstract{Surface tension (S) is due to the inward force
experienced by particles at the surface and usually gravitation
does not play an important role in this force.  But in compact
stars the gravitational force on the particles is very large and S
is found to depend not only on the interactions in the strange
quark matter, but also on the structure of the star, $i.e.$ on its
mass and radius.

~~~ Indeed, it has been claimed recently that 511 {\rm keV}
photons observed by the space probe INTEGRAL from the galactic
bulge may be due to $e^+~e^-$ annihilation, and their source may
be the positron cloud outside of an antiquark star. Such stars,
if they exist, may  also go a long way towards explaining away the
antibaryon deficit of the universe. For that to happen S must be
high enough to allow for survival of quark/antiquark stars born
in early stages of the formation of the universe.

~~~High value of S may also assist explanation of delayed
$\gamma~$ - ray burst after a supernova explosion, as conversion
from normal matter to strange matter takes place. The possibility
of some implications from formation of surface waves are also
discussed.

} \maketitle
\section{Introduction:}
~~~~For ordinary fluid, S is the property of the specific
interactions within and between the media forming an interface,
and gravitation does not play any significant role.

Neutron stars models cannot explain the properties of some
compact stars. These stars may be made entirely of deconfined
$u,~d,~s$ quark matter (\cite{Li99a, Li99b}) and may have been
formed in the early universe in a cosmic separation of phases
(Witten \cite{witten}). Alternatively they could be produced
after supernovae and show observed delayed gamma ray bursts as
signals of the transition from a normal matter remnant to a
strange one. Both models due to Alcock and Olinto \cite{3} and
Bombaci, Parenti and Vida\~na \cite{bom} (hence forth BPV) need
large values of S.

The MIT bag and the realistic strange star (ReSS) models are
considered. The maximum mass of the star and the corresponding
radius obtainable from the MIT bag does not significantly differ
from neutron star results. To get objects substantially more
compact than neutron stars the ReSS was invoked. The ReSS
incorporates chiral symmetry restoration (CSR), in a simple tree
level large $N_c$ model, with an interaction which contains both
asymptotic freedom (AF) and the confinement - deconfinement
mechanism (CDM) \cite{D98}. It can explain
\begin{enumerate}
\item the properties of compact stars like SAX J1808.8 or 4U
1728$-$34 (\cite{Li99a,Li99b}),
\item two quasi periodic peaks in X-ray power spectrum of some compact
stars (\cite{Li99b, bani}),
\item hours-long superbursts as diquark formation $-$ after
pairs are broken due to strong prolonged accretion seen in seven
stars \cite{mnras},
\item breathing mode absorption in 1E 1207$-$5209 \cite{mpla} and
excess emission bands in six other X-ray emitters \cite{MNRAS}.
\end{enumerate}

The SPI spectrometer on the INTEGRAL satellite has recently
detected a bright 511 {\rm keV} line in the $\gamma$ - ray
spectrum from the bulge of the galaxy with a spherically symmetric
distribution \cite{knodjean}. This has stimulated research in the
fundamental physics that describes cosmological dark matter.
Several authors (\cite{1}, Oaknin and Zhitnitsky \cite{2}) have
suggested that the line comes from positron annihilation,
possibly in strange stars formed of antiquarks. The suggestion of
cosmic separation of phases (Witten \cite{witten}) is that
strange stars (SS) can exist from the early universe and can
explain dark matter. A crucial condition for the survival of
these stars, laid down by Alcock and Olinto (\cite{3}), is a
large value of S.

One of the most interesting astrophysical observations of recent
times is the occurrence of supernova (SN) with remnants and
associated GRB with enormous energy, sometimes immediately
following the SN and sometimes delayed. In a recent paper, BPV
considered the possibility that GRBs may be delayed processes
following a supernova, in which ordinary matter is converted into
strange quark matter (SQM). Importantly, the scenario presented
by these authors also suggest that ``the existence of compact
stars with small radii (quark stars) does not exclude the
existence of compact stars with large radii (pure hadronic
stars), and vice versa." GRB delay is controlled by ``poorly
known" S according to these authors. Typical values used in the
literature range within 10 $-$ 50 ${\rm MeV~fm^{-2}}$ (Heiselberg
and Pethick \cite{heisel, ida, sato}). The variation in the
transition time from a normal matter remnant to a SQM star could
very well be due to a  variable S, which depends on the size of
the SQM star,  as we shall argue below.
\section{Surface tension : }
~~~We find from simple considerations that S varies from about 10
to 140 ${\rm MeV~fm^{-2}}$ depending on the star size. The  bag
model, which was historically the first one used to study SQM
gives values for S which is substantially smaller than the ReSS.

In addition to ReSS, in which the interquark potential is due to
Richardson, with a QCD scale parameter of 100 ${\rm MeV}$, we also
consider a ``new" ReSS  (\cite{mnj}) in which two scales are
chosen, one for asymptotic freedom ($\sim 100~{\rm MeV}$) and the
other for confinement ($\sim 350~{\rm MeV}$).

The surface at a radius $r$ is $4 \pi r^2$, the projection of a
quark on it is $\pi r_n^2$ - if it occupies a sphere of radius
$r_n~=~(1/{\pi n})^{1/3}$ on the average, where $n$ is the number
density. Heiselberg and Pethick (\cite{heisel}) suggested that the
quark scattering cross section $\pi r_n^2$ can be compared to
proton-proton scattering using the quark counting rule
$\sigma_{\rm pp} ~=~3~\sigma_{\rm qq}= ~3 ~\pi r_n^2$. Following
the above relation, we get the value of $\sigma_{\rm pp}$ to be
nearly $25~{\rm mb}$ which is quite reasonable as it is quoted to
be $25~{\rm mb}$ between  140 to 350 ${\rm MeV}$ (\cite{roy}).

This $r_n$ is comparable with interparticle separation
conventionally defined in nuclear physics for matter,
$r_0~=~(3/{4\pi n})^{1/3}$. $r_0$ and $r_n$ are compared in
Tables \ref{tb:ress2} and \ref{tb:bg2}.

Considering a SS as a huge drop of SQM, the pressure difference
across the surface can be expressed in terms of S :
 \be |\Delta p|_{r=R}~ =~\frac {2S}{R}. \label{eq:st} \ee
At the surface of a SS,
 \be \Delta p ~=~r_n~\frac{dp}{dr}
\ee equivalent to using a cut off ${\rm exp}(-r/r_n)$ to obtain
the pressure difference at the sharp boundary of the star. One
might recall that - unlike a neutron star which needs gravitation
to bind it - a SS is self-bound although that does not mean
gravitation plays a negligible part in keeping it more compact.

$\left|\frac{dp}{dr}\right|_{r=R}$ can be obtained by solving the
general relativistic equations for hydrostatic equilibrium (TOV
equation):
\be
\frac{dp}{dr}=\frac{-G[p(r)+\epsilon][m(r)+4\pi~r^3p(r)]}{r^2\left[1-\frac{2Gm(r)}{r}\right]}
\label{eq:tov} \ee

\begin{figure}
\resizebox{\hsize}{!}{\includegraphics{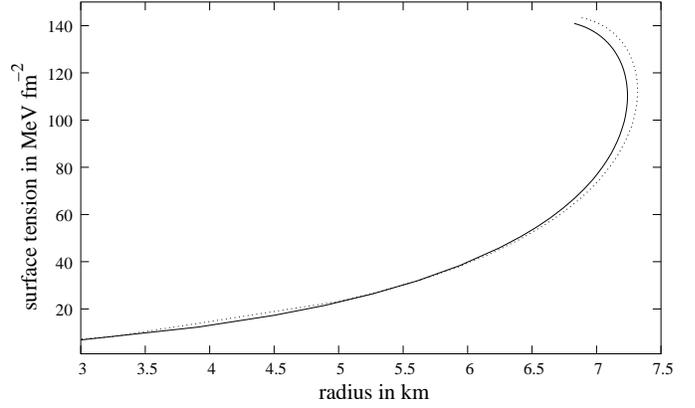}}
\caption{Variation of S with star size in ReSS and new ReSS. The
solid line is for the parameter set ReSS1 and the dotted line for
ReSSn1, given in table \ref{tb:ress2}. S is much larger for both
the cases showing that the strength of the interaction is much
larger in ReSS compared to the bag model (Fig.2).}
\label{fig:ress}
\end{figure}

\begin{figure}
\resizebox{\hsize}{!}{\includegraphics{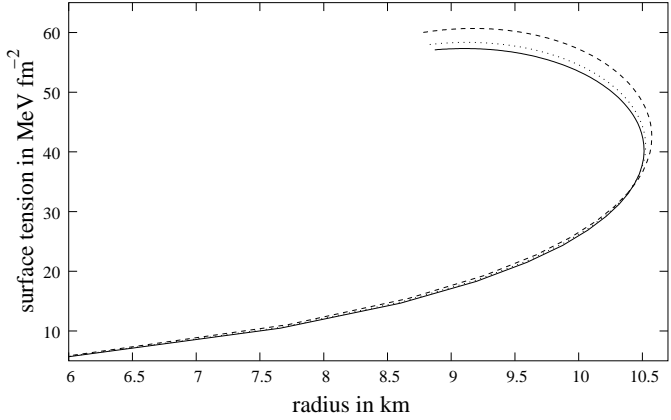}}
\caption{Variation of S with star size in Bag model for two sets
of parameters, showing that S increases appreciably only when
$\alpha_s$ is 0.5. The solid curve is for parameter set bg1, the
dotted for bg2 and the dashed for bg3, given in table
\ref{tb:bg2}.} \label{fig:bag}
\end{figure}

One gets $S\sim 140 ~{\rm MeV~fm^{-2}}$ for a SQM star of radius
6.95 ${\rm km}$ and mass of 1.42 ${\rm M_\odot}$ (see Fig
\ref{fig:ress}). The Newtonian limit (\cite{chiu}) leads to
$S\sim 36 ~{\rm MeV~ fm^{-2}}$ for a star of same radius. This
shows that the general relativistic effect is very important. It
can also be seen from Fig \ref{fig:ress} for radii 3 to 6 ${\rm
km}$ one gets the standard value of S varying from 10 to 50 ${\rm
MeV~ fm^{-2}}$, showing that our estimate of $r_n$ is realistic.
The bag model gives more moderate values for S which is
understood as being due to the less compact nature of the star
(see Fig \ref{fig:bag}).
\begin{table*}[htbp]
\caption{Results from  ReSS and new ReSS models. The central
density is $\rho_c$ and $\alpha_s$ is the strong coupling constant
that controls the Debye screening. The parameter $N$ controls the
CSR, through the quark mass equation, $m~ =~m_q ~~sech\left(
\frac{1}{N}\frac{n_B}{n_0}\right)$, where $n_B = (n_u + n_d +
n_s)/3$ is the baryon number density and $n _0$ is the normal
nuclear matter density. The strange quark mass
$m_s~=~m~+~150~{\rm MeV}~$.\label{tb:ress2} } \centering
% \vskip .5cm
\begin{tabular}{ c c c c c c c c c c c }
\hline \hline
&\multicolumn{3}{c}{Parameters}&\multicolumn{7}{c}{Maximum mass
and corresponding results}
\\
\cline{1-4} \cline{5-11} EOS &N&$m_q$&$\alpha_s$ &Mass &$\rho_c$ &
Radius &$\left|\frac{dp}{dr}\right|_{r=R}$ &$r_n$& $r_0$ & S \\

&&${\rm MeV}$& &${\rm M_\odot}$ &$10^{14}~{\rm gm/cm^{3}}$   &
${\rm km}$ & ${\rm MeV/fm^{3}/km}$ &${\rm fm}$&${\rm fm}$ & ${\rm
MeV^3}$
\\ \hline
ReSS &3.0&310&0.2  &1.44&46.85&7.06  &73.87 &0.51&0.47&$
(173.5)^3$\\
  ReSSn&3.0&325&0.65& 1.47 &45.49& 7.14&73.05 &0.52&0.47 &$(173.9)^3$\\ \hline
\end{tabular}
\end{table*}
\begin{table*}
\caption{Parameters and results for bag models. The masses $m_u~ =
~0~=~ m_d$ and $m_s~=~150~{\rm MeV}~$. \label{tb:bg2}}
 \centering
%\vskip .2cm
\begin{tabular}{c c c c  c c c c c c }
\hline\hline
&\multicolumn{2}{c}{Parameters}&\multicolumn{7}{c}{Maximum mass
and corresponding results}
\\ \cline{1-4}\cline{5-10} EOS&B&$\alpha_s$ & Mass &$\rho_c$  & Radius
&$\left|\frac{dp}{dr}\right|_{r=R}$  &$r_n$&$r_0$& S\\
 &${\rm MeV/fm^3}$&& ${\rm M_\odot}$ &$10^{14}~{\rm gm/cm^{3}}$  & ${\rm km}$
&${\rm MeV/fm^{3}/km}$ &${\rm fm}$& ${\rm fm}$ &${\rm MeV^3}$\\
\hline
 bg1 &60&0.0 &1.81&23.17&10.07  &14.75 & 0.71&0.64 &$(123.1)^3$ \\
  bg2&60&0.17 &1.82&23.18&10.08& 14.93&0.71&0.65 &$(123.1)^3$ \\
  bg3&60&0.50 &1.85&23.20&10.12&15.32&0.72&0.66 &$(125.6)^3$\\
  bg4&60&0.17 &1.75 &24.85&9.76&15.61&0.71&0.65 &$(126.1)^3$ \\
 bg5 &75&0.17 &1.64&28.81&9.06 &20.71 &0.68 &0.61 &$(130.9)^3$\\
 \hline
\end{tabular}
\end{table*}
\noindent for both tables the strange quark has an additional mass
of 150 ${\rm MeV}$.

In the graphs, we have chosen ${\rm MeV ~fm^{-2}}$ as the unit of
S ; while in the tables ${\rm MeV^3}$ is the chosen unit for
surface tension as given by Alcock and Olinto (\cite{3}) to
facilitate comparison with them. They already indicated that
lumps of SQM might have survived dissolution into hadrons in the
early universe if S is around $(178~{\rm MeV)^3}$. The S
calculated from ReSS and new ReSS gives results of this order
while that from the bag model is low.

\section{Other implications:}
Type I X-ray bursts from seven LMXBs have been well studied
through RXTE. Often a peak is observed in the power density
spectrum (PDS) at a frequency which is taken as the compact star's
spin frequency (or half of the spin frequency). But astonishingly,
the frequency is not constant throughout the burst. It shifts
towards a higher value. There are several possible explanation
such as photospheric radius expansion \cite{stroh97}, Rossby mode
vibration \cite{heyl} etc. We try to explain it as the onset of a
surface wave as the burst proceeds and the shift of the peak is a
result of the coupling of the surface wave to the burst power.

Applying Bernoulli's theorem to the crest and trough of the
sinusoidal wave, we get the equation :
\begin{eqnarray}
\nonumber
\frac{GM(R+h)}{{(R+h)}^2-{R_s}^2}~+~\frac{1}{2}{\left({\rm
V}+\frac{2\pi
b}{T}\right)}^2~-~\frac{Sc^2}{b\epsilon} \\
=\frac{GM(R-h)}{{(R-h)}^2-{R_s}^2}~+~\frac{1}{2}{\left({\rm
V}-\frac{2\pi b}{T}\right)}^2~+~\frac{Sc^2}{b\epsilon}
\label{eq:ber1}
\end{eqnarray}
where S, M, R, $R_s$, c are the surface tension, mass, radius, the
Schwarzchild radius of the star and velocity of light. $\epsilon$
is the energy density at the star surface. h, T, V are
respectively the amplitude, time period and velocity of the wave.
The radius of curvature at crest and trough, b, is given by :\be
b=~\frac{{(1+{(dy/dx)^2)}}^{3/2}}{d^2y/dx^2}~=~\frac{{\lambda}^2}{4{\pi}^2h}~=~\frac{{\rm
V}^2T^2}{4 \pi^2 h }\ee

With the correction term $2{\pi}b/T$ in the velocity, V, due to
circular motion of the particle, the eqn (\ref{eq:ber1}) reduces
to:
\begin{eqnarray}
\nonumber \frac{{\pi}h {\rm V}^2{R_s
c^2}}{2T}\left[\frac{(R+h)}{{(R+h)}^2-{R_s}^2}~-\frac{(R-h)}{{(R-h)}^2-{R_s}^2}\right]\\=-\left[{\rm
V}^5~+\frac{8{\pi}^3h^2Sc^2}{\epsilon
T^3}\right]~,\label{eq:ber2}\end{eqnarray}

\begin{figure}
\resizebox{\hsize}{!}{\includegraphics{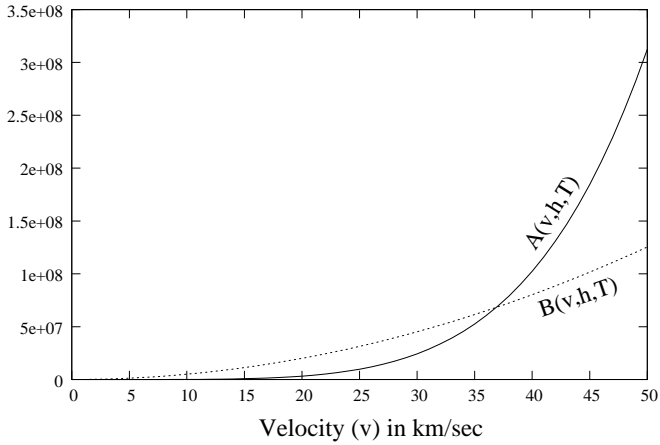}}
\caption{Solution of equation \ref{eq:ber2} for $V=36.89~{\rm
km/sec}$ and $T~=~2/3~{\rm sec}$.} \label{fig:poly}
\end{figure}

We solve eqn (\ref{eq:ber2}) graphically; both R.H.S and L.H.S.
are functions of V, $h$ and $T$ - they can be denoted by A(V, $h$,
$T$) and B(V, $h$, $T$). Different values of V can be obtained for
combinations of $h$ and $T$. Figure \ref{fig:poly} shows one such
solution with $T~=~ 2/3$ ${\rm sec}$, $V~=~36.89~{\rm km/sec}$
giving $h~=~64.76~{\rm cm}$.

\begin{figure}
\resizebox{\hsize}{!}{\includegraphics{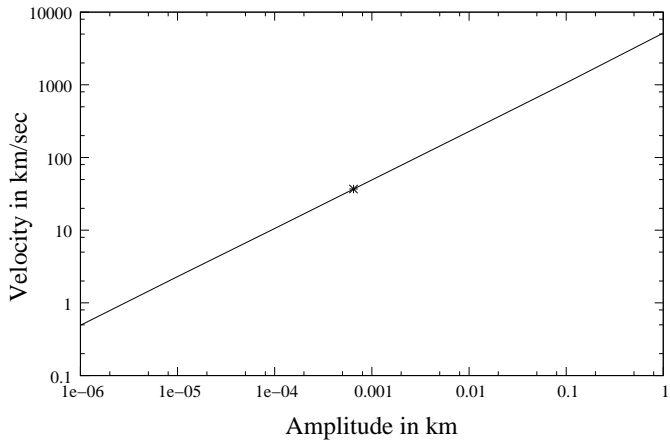}}
\caption{Variation of V with $h$ for a wide range of values of
$h$, keeping $T$ fixed at $2/3~{\rm sec}$. The $\ast$ point
denotes the set of V, h required to explain the $\Delta$f,
observed in 4U 1728$-$34.} \label{fig:varh_f1.5}
\end{figure}

\begin{figure}
\resizebox{\hsize}{!}{\includegraphics{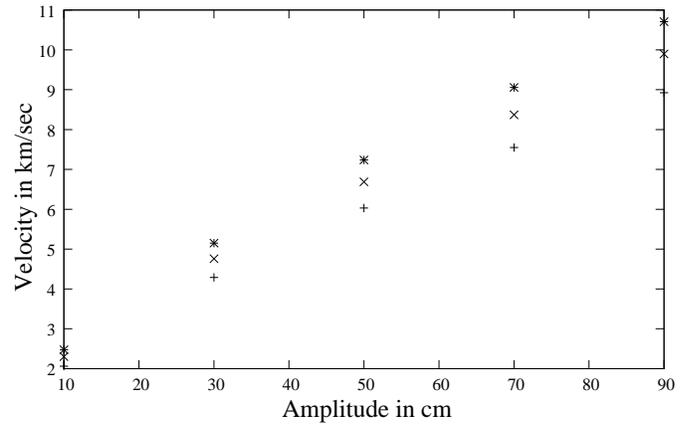}}
\caption{Variation of V with $h$ for different $T$. The {$\ast$}
's are for $T~=~0.52632~{\rm sec}$ ($\Delta$f=1.9), the
{$\times$} 's are for $T~=~2/3~ {\rm sec}$ ($\Delta$f= 1.5) and
the {+} 's are for $T~=~0.90909~{\rm sec}$ ($\Delta$f=1.1).}
\label{fig:varhdift}
\end{figure}

To fix V and $T$ we turn to the burst from the compact star 4U
1728$-$34 (\cite[]{stroha} \cite{strohb}), the rise time of the
burst is 0.6 seconds and the frequency shift ($\Delta$f) observed
in the maximum of PDS is 1.5 {\rm Hz}. Assuming the surface wave
originates at the burst spot and travels over the entire surface
of the star within the burst rise time - its velocity is 36.89
${\rm km/sec}$ for a star with half the circumference of $7.05
\pi$ ${\rm km}$. We take the time period of the wave as the
inverse of the frequency shift $i.e.,~T~=~2/3~{\rm sec}$. From
these values of V and $T$, we get $h~=~64.75~{\rm cm}$ from
Figure \ref{fig:poly}. These values are therefore consistent. For
theoretical interest, in figure \ref{fig:varh_f1.5} we plot the
velocity of the wave for a wide range of the amplitude, keeping
$T$ fixed at 2/3 ${\rm sec}$. In figure \ref{fig:varhdift}, we
plot the variation of V with $h$ for different values of $T$,
keeping $h$ in order of ${\rm cm}$ and $T$ also near to $2/3~{\rm
sec}$.

We thank the anonymous referee for reminding us that a SS has an
electron cloud outside it and this leads to a electrostatic field
which may be as strong as $10^{17}~ {\rm Vcm^{-1}}$ decreasing to
$10^{11}$ as one goes out radially by one $\AA$ (see \cite{xu}
for a review and update ). This may affect the above estimate and
will be considered elsewhere. The accretion rate on stars vary a
lot, the maximum being $10^{-5}~{\rm M_\odot/yr}$ and this has to
be taken into account. Finally, the surface wave may cause
gravitational radiation which should be observable in the future.
The amplitude of the surface wave that we have estimated is of
order $50 \sim 100~ {\rm cm}$, as such it will not cause serious
deformation of the shape of a $R~=~7 ~{\rm km}$ star.
\section{Conclusions and summary:} The high values of S found for the ReSS
and new ReSS keep open the possibility of cosmic separation in the
early universe, which in turn may explain the baryon-antibaryon
asymmetry and the dark matter problem according to Oaknin and
Zhitnitsky (\cite{2}). The large value of surface tension could
also help in explaining the variable delay time between a SN and a
GRB in the scenario of BPV. It is amusing to find that S in dense
systems do after all depend on gravitation. With our estimated
value of S, we can also explain the frequency shift in the PDS of
type I X-ray burst.

\begin{acknowledgements}
M. B. checked the results of 4U 1728$-$34 at TIFR thanks to the
hospitality of Dr. Biswajit Paul and thanks Uddipan Mukherjee for
help.
\end{acknowledgements}


\begin{thebibliography}{}
\bibitem[1989]{3} Alcock, C. \& Olinto, A., \prd, 1989, 39, 1233.

\bibitem[Bagchi, Ray, Dey and Dey 2005]{mnj} Bagchi, M., Ray, S., Dey, M. \& Dey, J.,
(communicated to A\&A. ).

\bibitem[Boehm, Hooper, Silk, Casse 2004]{1}  Boehm, C., Hooper, D., Silk, J., Casse, M. \& Paul, J.
Phys. Rev. Lett. 2004, 92, 101301.

\bibitem[2004]{bom} Bombaci, I., Parenti, I. \& Vida\~na, I., \apj,
2004, 614, 314.

\bibitem[Chiu 1968]{chiu} Chiu, H. Y., Stellar Physics (Blaisdell
Publishing Company, 1968), p. 70.

\bibitem[(Dey, Bombaci, Dey, Ray and Samanta 1998)]{D98} Dey, M., Bombaci, I., Dey, J., Ray, S.
\& Samanta, B. C., Phys.  Lett. B 1998, 438, 123; Addendum 1999,
447, 352; Erratum B 1999 467, 303; Indian J. Phys. B 1999, 73,
377.

\bibitem[1993]{heisel} Heiselberg, H. \& Pethick, C. J.,
\prd, 1993, 48, 2916.

\bibitem[(Heyl 2004)]{heyl} Heyl, J. S., \apj, 2004, 600, 939.

\bibitem[Iida and Sato 1997]{ida} Iida, K. \& Sato, K., Prog. Theo. Phys., 1997, 98, 277.

\bibitem[1998]{sato} Iida, K. \& Sato, K., \prc, 1998, 58, 2538.


\bibitem[(Knodleseder, Lonjou, Jean $et~al$ 2003)]{knodjean} Knodleseder, J.,
Lonjou, V., Jean, P. $et~al.$, Astron. \& Astrophys., 2003, 411,
L457.

\bibitem[Li, Bombaci, Dey, Dey and Heuvel 1999]{Li99a} Li, X., Bombaci, I., Dey, M., Dey \& J. and van den
Heuvel, E. P. J., \prl, 1999, 83, 3776.

\bibitem[Li, Ray, Dey, Dey and Bombaci 1999]{Li99b} Li, X., Ray, S., Dey, J., Dey, M. \& Bombaci, I.
\apj, 1999, 527, L51.

\bibitem[Mukhopadhyay, Ray, Dey and Dey 2003]{bani} Mukhopadhyay, B., Ray, S., Dey, J. \& Dey,
M., \apj, 2003, 584, L83.

\bibitem[2004]{2} Oaknin, D. H., \& Zhitnitsky, A. R., Phys. Rev. Lett., 94, 2005,
101301.

\bibitem[(Ray, Dey, Dey and Bhowmick 2004)]{MNRAS} Ray, S., Dey, J., Dey, M. \&
Bhowmick, S., Mon. Not. Roy. Astron. Soc., 2004, 353, 825.

\bibitem[Roy and Nigam]{roy} Roy, R. R., Nigam, B. P., ``Nuclear
Physics: Theory and Experiment" (New Age International Publishers
1996, New Delhi), p. 89.

\bibitem[(Sinha, Dey, Ray and Dey 2002)]{mnras} Sinha, M., Dey, M., Ray S., \& Dey J., Mon. Not. Roy.
Astron. Soc. 2002, 337, 1368.

\bibitem[(Sinha, Dey, Dey, Ray and Bhowmick 2003)]{mpla}  Sinha, M., Dey, J., Dey, M., Ray, S. \& Bhowmick, S.
Mod. Phys. Lett. A, 2003, 18, 661.

\bibitem[Strohmayer, Zhang, Swank and Titarchuk 1996]{stroha} Strohmayer, T. E., Zhang, W., Swank,  J. H.,
Titarchuk, L. \& Day, C., 1996, \apj, 469, L9.

\bibitem[(Strohmayer, Jahoda, Giles and Lee 1997)]{stroh97} Strohmayer, T. E., Jahoda, K., Giles,
A. B. \& Lee, U., 1997, \apj  486, 355.


\bibitem[Strohmayer, Zhang, Swank and White 1998]{strohb}Strohmayer, T. E., Zhang, W.,  Swank, J. H.
\& White, N. E., 1998, \apj, 498, L135.

\bibitem[1984]{witten} Witten, E., \prd, 1984, 30, 272.

\bibitem[Xu, Zhang and Qiao 2001]{xu} Xu, R. X., Zhang, B. and Qiao, G. J., Astropart. Phys., 2001, 15, 101.

\end{thebibliography}
\end{document}